\documentclass[prd, twocolumn, superscriptaddress, preprintnumbers, fixfloat, nofootinbib]{revtex4-1}

\newcommand{\redtext}[1]{\textcolor{black}{#1}}
\newcommand{\hrm}[1]{\textcolor{black}{#1}}

\newcommand{\bluetext}[1]{\textcolor{black}{#1}}

\newcommand{\dcc}{LIGO-P2000174}

\usepackage[dvipsnames]{xcolor}
\usepackage{amsmath}
\usepackage[colorlinks=true, linkcolor=blue, citecolor=green]{hyperref}
\usepackage{graphicx}
\usepackage{subfigure}
\usepackage{esvect}

\graphicspath{{figures/}}
\begin{document}

\title{Measuring angular $N$-point correlations of binary \hrm{black hole} merger gravitational-wave events with hierarchical Bayesian inference.}

\author{Sharan Banagiri}
\email{banag002@umn.edu}
\affiliation{School of Physics and Astronomy, University of Minnesota, Minneapolis, MN 55455, USA}

\author{Vuk Mandic}
\affiliation{School of Physics and Astronomy, University of Minnesota, Minneapolis, MN 55455, USA}

\author{Claudia Scarlata}
\affiliation{School of Physics and Astronomy, University of Minnesota, Minneapolis, MN 55455, USA}

\author{Kate Z. Yang}
\affiliation{School of Physics and Astronomy, University of Minnesota, Minneapolis, MN 55455, USA}

\date{\today}%

\begin{abstract}
    Advanced LIGO and Virgo have detected ten binary black hole mergers by the end of their second observing run. These mergers have already allowed constraints to be placed on the population distribution of \hrm{black holes} in the Universe, which will only improve with more detections and increasing sensitivity of the detectors. In this paper we develop techniques to measure the angular distribution of \hrm{black hole} mergers by measuring their statistical $N$-point correlations through hierarchical Bayesian inference. We apply it to the special case of two-point angular correlations using a Legendre polynomial basis on the sky. Building on the mixture model formalism introduced in Ref.~\cite{Smith:2017vfk} we show how one can measure two-point correlations with no threshold on significance, allowing us to target the ensemble of sub-threshold binary black hole mergers not resolvable with the current generation of ground based detectors. We also show how one can use these methods to correlate gravitational waves with other probes of large scale angular structure like galaxy counts, and validate both techniques through simulations.  
\end{abstract}

\maketitle

\section{Introduction}
\label{Sec:Intro}
The direct detections of gravitational waves (GW) by the advanced Laser Interferometer Gravitational-wave Observatory (aLIGO) and advanced Virgo (aVirgo) detectors~\cite{LVC:150914, LVC:170817, LVC:O2Catalog} have given us a new tool to probe the Universe. GW can carry astrophysical and cosmological information not accessible through electromagnetic observations. This is especially true for information about \hrm{black hole} mergers which leave no electromagnetic trace \redtext{\footnote{However there have been a few claims of potential electromagnetic counterparts to binary black hole mergers, see~\cite{Connaughton:2016umz, Graham:2020gwr}}}. The aLIGO and aVirgo detectors have detected ten binary black hole (BBH) mergers~\cite{LVC:O2Catalog} in the first two observing runs (O1 and O2), with many more candidate events in the recently completed third observing run (O3)~\cite{gracedb}. These detections have allowed for constraining the mass, spin and redshift distributions of BBH progenitors, along with measuring the rate of mergers in the local Universe~\cite{LVC:O2Rates&Pop}. Additional events have also been claimed by groups analyzing publicly available data from O1 and O2~\cite{Nitz:2019hdf,Venumadhav:2019lyq,Venumadhav:2019tad}. \redtext{Recently there have also been studies measuring the angular distributions of the BBH merger events in the published GWTC-1 catalog from O1 and O2~\cite{Stiskalek:2020wbj, Payne:2020pmc}.} 

With the current generation of ground based GW detectors, there are many more such mergers \bluetext{of stellar origin} which are individually unresolvable. Finding the signal from this ensemble of unresolved mergers has traditionally been a key target of stochastic \redtext{GW} searches~\cite{LVC:O2stochatic, Allen:1997ad, Romano:2016dpx} that employ \hrm{cross correlation} between detectors to detect an astrophysical background. Multiple analyses have been developed to measure anisotropies in the stochastic background as applied to ground based GW detector data~\cite{Thrane:2009aa, TheLIGOScientific:2016xzw, LIGOScientific:2019gaw, Renzini:2019vmt, Ain:2018zvo}. In recent years, there have also been theoretical predictions about the anisotropic properties of astrophysical BBH backgrounds~\cite{Cusin:2018rsq, Jenkins:2018kxc,Jenkins:2018uac, Bertacca:2019fnt, Cusin:2017fwz, Cusin:2017mjm, Cusin:2019jpv, Pitrou:2019rjz, Contaldi:2016koz} which suggest that the lowest order multipoles ---  traditionally thought to be the ones stochastic searches are most sensitive to --- are at least an order of magnitude smaller than the monopole, placing them beyond the reach of current generation detectors using standard \hrm{cross correlation} based methods. Moreover, because of the relatively small number of \redtext{BBH merger} events that occur in an observation time, shot noise effects from statistical Poisson error can dominate over the astrophysical contribution to the higher order multipoles, making their discovery and measurement further difficult~\cite{Jenkins:2019uzp, Jenkins:2019nks}.

In this paper we develop methods to probe the statistical properties of the angular distribution of the ensemble of BBH mergers, folding in the discrete nature of events. We first construct ways to measure the angular $N$-point correlations of the background through hierarchical Bayesian inference with a focus on the special case of two-point correlations, and apply it to simulations using the mixture model framework developed by Smith and Thrane in Ref.~\cite{Smith:2017vfk}. 

The mixture model approach places no thresholds on the signal-to-noise ratio (SNR) of the events and can jointly draw inferences from well resolved events and events in the astrophysical background (the so called sub-threshold events) accounting for their discrete nature, and removing the somewhat artificial distinction between classical compact binary coalescence (CBC) signals and stochastic \redtext{GW} backgrounds from CBC sources. The mixture model framework promises in general to be much more sensitive than the cross-correlation search in probing the astrophysical background since it looks for very specific signals based on CBC waveform models, and with the addition of hierarchical Bayesian inference can also be used to estimate properties of the population of BBH systems~\cite{Smith:2020lkj}  in addition to the rate of their mergers. This extra sensitivity can be promising in attempting to detect the angular structure of the ensemble of binary mergers.

The rest of the paper is organized as follows. In Sec.~\ref{Sec:NptCprr} we define the $N$-point correlation function and using that as a correlated prior we write down an expression for the Bayesian signal evidence. In Sec.~\ref{Sec:2ptCorr} we use this to write the Bayesian evidence for two-point correlations between BBH mergers. We also show how a similar expression for evidence can be written for \hrm{cross correlating} BBH mergers with a different tracer of the large scale structure like galaxy counts. Building upon the mixture model from Ref.~\cite{Smith:2017vfk}, in Sec.~\ref{Sec:TBS} we show how the methods from the previous sections can be used for Bayesian inference of anisotropies in an ensemble of BBH mergers. Sec.~\ref{Sec:Simulations} describes the simulations and presents results for the two-point correlation method for both BBH-BBH and BBH-galaxy count correlations, followed by a discussion in Sec.~\ref{Sec:DQ} of potential data quality issues when these techniques are applied to real data.

\section{BBH $N$-point Correlation}
\label{Sec:NptCprr}

Statistical $N$-point correlation functions are traditionally used in measuring the clustering of galaxies~\cite{Peebles:1981}, with the most important statistic perhaps being the two-point correlation function. \bluetext{Since BBH mergers are discrete in nature, we can correlate multiple observed events to measure the statistical correlations of the background, and probe the angular structure of the ensemble of mergers and their progenitors}. The tool of choice for population inference is hierarchical Bayesian modeling, in which a suitably modeled prior contains the population parameters to be measured. We point to Ref.~\cite{thrane_talbot_2019} for an overview of Bayesian inference in the field of GW analysis, including hierarchical inference, \bluetext{and to Ref.~\cite{LIGOScientific:2019hgc} for a guide to aLIGO-aVirgo noise and data.} 

Given a GW data segment $d$, the Bayesian evidence that it contains a GW signal is given by:

\begin{equation}
	      \mathcal{Z}^0= \int d \vec{\lambda} \, \mathcal{L} (d | \vec{\lambda}) \, \pi^0 (\vec{\lambda}),
	      \label{Eq:simple_evidence}
\end{equation}
where $\vec{\lambda}$ includes all relevant parameters needed for describing BBH merger waveforms like masses, spins, distance, sky position etc. The term $\mathcal{L} (d | \vec{\lambda})$ is the standard likelihood used in GW searches based on the noise statistics of colored, \hrm{frequency domain} Gaussian noise~\cite{Veitch:2014wba, Ashton:2018jfp}, and $\pi^0 (\vec{\lambda})$ are some standard fiducial priors on $\vec{\lambda}$. Note that the parameters describing the sky position $\hat{\Omega} = ( \theta, \phi)$ are part of $\vec{\lambda}$. 

\bluetext{Measuring the $N$-th statistical correlation requires $N$ data segments each containing a signal}. Firstly, under the standard assumption that they are independent of each other and using the fiducial priors, the evidence for the hypothesis that all of them contain a BBH signal~\footnote{We will assume that the probability for a data segment to contain more than single BBH merger is negligible. Similarly the probability of a signal cutting across segments is assumed to be negligible. The duration $T$ of the data segment will have to be chosen to ensure that this is a good assumption.} follows from Eq.~\ref{Eq:simple_evidence} as, 

\begin{equation}
      \mathcal{Z}^0_N = \int \prod_i^N \left [d \vec{\lambda}_i \, \mathcal{L} (d_i | \vec{\lambda}_i) \, \pi^0 (\vec{\lambda}_i) \right ],  
      \label{Eq:NsegEvidence}
\end{equation}
where $i$ is an index over data segments. To measure correlations in the population distribution we can use a joint correlated prior between multiple segments. To write the evidence for the $N$-point angular correlations hypothesis we replace the fiducial angular prior in Eq.~\ref{Eq:NsegEvidence} with a prior correlated across the $N$ data segments, while retaining the fiducial priors on all other parameters, viz~\footnote{In this paper we have chosen to describe $\mathcal{Z}^c_N (\vec{\Lambda})$ as the Bayesian evidence. An alternate terminology used in many papers in the literature is to call it the posterior on $\vec{\Lambda}$ marginalized over all the intrinsic parameters of the mergers.}:

\begin{equation}
      \mathcal{Z}^c_N (\vec{\Lambda})= \int \left[\zeta^N ( \{ \hat{\Omega}_{j} \}  |  \vec{\Lambda}) \,\prod_{i=1}^N \left (d \vec{\lambda}_i \,  \frac{\mathcal{L} (d_i | \vec{\lambda}_i)  \pi^0 (\vec{\lambda}_i)}{\pi^0 (\hat{\Omega}_i)} \right ) \right],  
      \label{Eq:NpointEvidence}
\end{equation}
where $\pi^0 (\hat{\Omega}_i)$ are the fiducial \textit{uncorrelated} priors on angular parameters of each segment. The term $\{ \hat{\Omega}_{j} \}$ refers to the set of all the directional parameters for the $N$ segments, with the $N$-point correlation function $\zeta^N ( \{ \hat{\Omega}_{i} \}  |  \vec{\Lambda})$ being the joint prior on them with some \hrm{hyperparameters} $\vec{\Lambda}$. The \hrm{hyperparameters} describe the population distribution of the mergers, in this case the angular distribution. 
We interpret the correlation function as the probability density function of $N$ objects being at the angular positions of $\{ \hat{\Omega}_{j} \}$ depending on the choice of  $\vec{\Lambda}$~\cite{Peebles:1981}. 

\bluetext{ It is important to note that the individual BBH mergers are still treated as independent events -- as they must be since they spatially and temporally separated -- which allows the likelihoods in Eq.~\ref{Eq:NpointEvidence} to be multiplied with each other. The statistical correlations only describe the population distribution of the BBH mergers.}

The posterior distributions are usually calculated by the means of nested or Markov chain Monte Carlo samplers. The Eq.~\ref{Eq:NpointEvidence} above casts the problem as one of joint inference of both \hrm{hyperparameters} $\vec{\Lambda}$ and parameters of the BBH merger $\vec{\lambda}_i$. However, generally it is not necessary to redo the sampling at every point in the \hrm{hyperparameter} space. Instead the posterior samples generated while using the fiducial choice of priors on the BBH parameters $\vec{\lambda}_i$ can be \textit{recycled} to provide inferences in the \hrm{hyperparameter} space~\cite{thrane_talbot_2019}. To see this in the case of the directional parameters, we first use Bayes theorem on Eq.~\ref{Eq:NpointEvidence} to get: 

\begin{equation}
      \mathcal{Z}^c_N (\vec{\Lambda})=  \int \left[\mathcal{Z}^0_N \,\zeta^N ( \{ \hat{\Omega}_{j} \}  |  \vec{\Lambda}) \,\prod_{i=1}^N \left (d \vec{\lambda}_i \,  \frac{P^0 (\vec{\lambda}_i | d_i)}{\pi^0 (\hat{\Omega}_i)} \right ) \right], 
\end{equation}
where $ P^0 (\vec{\lambda}_i| d_i )$ is the fiducial posterior for segment $i$ obtained using prior $\pi^0 (\vec{\lambda}_i)$. The evidence $\mathcal{Z}^0_N$ (defined in Eq.~\ref{Eq:NsegEvidence})  comes up as the normalization factor when using Bayes theorem and can be pulled out of the integral. We then marginalize over all the non-directional parameters to get:

\begin{equation}
	      \mathcal{Z}^c_N (\vec{\Lambda})= \mathcal{Z}^0_N \int \left[\zeta^N ( \{ \hat{\Omega}_{j} \}  |  \vec{\Lambda})\prod_i^N \left (d^2 \hat{\Omega}_i \,  \frac{P^0 (\hat{\Omega}_i | d_i )}{\pi^0 (\hat{\Omega}_i)}\right ) \right].
\end{equation}
 The integrand is then just the expectation value of the ratio of the new correlated prior and the fiducial angular priors. When we have posterior samples rather than a continuous measurement of $ P^0 (\hat{\Omega}_i | d_i )$ this can be approximated as, 

\begin{equation}
    \mathcal{Z}^c_N (\vec{\Lambda}) \approx  \mathcal{Z}^0_N  \sum \, \frac{\zeta ( \{\hat{\Omega}^{m^i}_{i}\} \, | \, \vec{\Lambda})}{ \prod_i  M_i \, \pi^0 (\hat{\Omega}_i^{m^i})}  .
    \label{Eq:Npt_discrete_evidence}
\end{equation}
Here $i$ is an index over segments while ${m^i} = 1 ... M_i$ is an index over the posterior samples of that segment, and $M_i$ is the number of posterior samples for that segment. The summation is over all possible $N$-point correlations between posterior samples across segments the total number of which is given by $\prod_i M_i$. For the directional parameters $\hat{\Omega}_i$, the fiducial angular priors are usually taken to be isotropic for each segment i.e $\pi^0 (\hat{\Omega}_i)  = 1/{4\pi}$.

\section{Two-Point correlation}
\label{Sec:2ptCorr}

\subsection{BBH Two-Point Correlation}
\label{Sec:BBH-2pt}

When probing large scale cosmological structure one is usually more interested in the statistical properties of the distribution rather than the specific realization in our universe. Theoretical models also only predict the statistical properties. Statistical isotropy is a standard assumption made when measuring cosmological correlations, which often simplifies Bayesian searches by reducing the number of parameters needed for modeling structures of a given angular scale. With these assumptions, measuring the two-point correlation of the BBH background is the most interesting and the simplest case of $N$-point correlations. Two-point correlation- based priors could also be used in a straightforward manner for directly correlating GW data with EM probes of structure. Hence in this section and the rest of the paper we apply the $N$-point correlation formalism to the specific case of $N = 2$. 

 Assuming statistical isotropy and homogeneity we define the two-point correlation function as the probability density of two BBH mergers $i$ and $j$ being at an angular separation $\Delta_{ij}$~\cite{Peebles:1981, peacock:1998}. Under these assumptions, the correlation function $\zeta$ can be expanded in the basis of Legendre polynomials $\mathcal{P}_{\ell}(  \Delta_{ij})$ with coefficients $\{C_\ell\}$ as parameters:

\begin{equation}
    \zeta ( \Delta^{mn}_{ij} \, | \, \{ C_\ell \} ) = \frac{1}{(4 \pi)^2} \sum_{\ell} (2 \ell + 1) \,C_\ell \, \mathcal{P}_{\ell} (\cos (\Delta^{mn}_{ij})),
    \label{Eq:Legandre_decomposition}
\end{equation}
where the factor of $(4 \pi)^2$ is required to normalize $\zeta$ as a probability distribution function. The convention we adopt here is to define $\zeta $ to be the complete two-point correlation function instead of the over and under densities as is usually done in galaxy number count analyses. Our definition can be converted into the latter by simply writing the isotropic (monopole) term separately. If  $\hat{\Omega}_i = (\theta_i, \phi_i)$ and $\hat{\Omega}_j = (\theta_j, \phi_j)$ are the coordinates of the BBH merger events, the angular separation between them on the \hrm{two sphere}  $ \Delta_{ij} $ is given by:

\begin{equation}
   \cos(\Delta_{ij}) = \sin \theta_i \sin \theta_j \cos (\phi_i - \phi_j) +  \cos \theta_i \cos \theta_j.
\end{equation}
 The joint evidence $\mathcal{Z}^c_{ij}$ for the two-point correlation hypothesis as a function of $\{C_{\ell}\}$ then follows from Eq.~\ref{Eq:NpointEvidence}:

\begin{equation}
\begin{split}
    \mathcal{Z}^c_{ij}(\{C_\ell \}) = & \int d\vec{\lambda}_i \, d \vec{\lambda}_j \, \pi^0 (\vec{\lambda}_i)  \, \pi^0 (\vec{\lambda}_j)  \\
    & \times \frac{\mathcal{L} (d_i | \vec{\lambda}_i) \mathcal{L} (d_j | \vec{\lambda}_j)}{\pi^0 (\hat{\Omega}_i)\pi^0 (\hat{\Omega}_j) } \, \zeta ( \Delta_{ij} | \{C_\ell \}), 
\end{split}
\end{equation}
with the equivalent of Eq.~\ref{Eq:Npt_discrete_evidence} given by, 

\begin{equation}
 \mathcal{Z}^c_{ij} (\{C_\ell \}) \approx  \frac{\mathcal{Z}^0_i   \mathcal{Z}^0_j}{M_i M_j} \sum_m \sum_n \, \frac{\zeta ( \Delta^{mn}_{ij} \, | \, \{C_\ell \})}{\pi^0 (\hat{\Omega}_i^m) \pi^0 (\hat{\Omega}_j^n)}.
 \label{Eq:2pt_corr_evidence}
\end{equation}
Here $\mathcal{Z}^0_{i}$ and $\mathcal{Z}^0_{j}$  are the signal evidences using the fiducial angular priors for event $i$ and $j$. The term $ \Delta^{mn}_{ij}$ is the angular separation between the $m$-th sample and the $n$-th sample in the posteriors of the $i$ and $j$ \hrm{data segments} respectively. 
Finally using Eq.~\ref{Eq:Legandre_decomposition} and the isotropic value for $\pi_o(\hat{\Omega}_i) = 1/4\pi$ we get,

\begin{equation}
 \mathcal{Z}^c_{ij} (\{C_\ell \}) \approx  \frac{\mathcal{Z}^0_i   \mathcal{Z}^0_j}{M_i M_j}  \sum_{m, n}  \sum_{\ell}  (2 \ell + 1) \,C_\ell \, \mathcal{P}_{\ell} \left (\cos (\Delta^{mn}_{ij} ) \right).
 \label{Eq:Legdenre_2pt_decomposition}
\end{equation}

\subsection{BBH - Galaxy two-point correlation}
\label{Sec:BBH-Gal-2pt}

If the progenitors of BBH mergers are \hrm{black holes} of stellar origin, we expect that their angular distribution will follow that of the large scale structure on the sky. \hrm{Cross correlating} this distribution with other tracers of structure like galaxy counts will allow us to probe this common matter distribution, and also test theories of structure and evolution. In this section we show how a two-point correlation evidence can also be written for measuring \hrm{cross correlations} between GW and galaxy distribution as seen by surveys like SDSS~\cite{SDSS1_2017}. The theoretical prediction of the \hrm{cross correlation} will depend on details of redshift evolution of the star formation rate among other things but we can again write it in a fairly model independent way as an expansion in Legendre polynomials, assuming statistical isotropy. 

In the case of \hrm{cross correlation} we now define the two-point correlation function $\zeta_{ij} (  \Delta_{ij}  | \{C_\ell \})$ as the probability of having a BBH merger $i$ at an angular separation of $ \Delta_{ij} $ from a galaxy $j$ \footnote{ There are two distinct but related questions we can ask. The first is the probability of having a BBH event at an angular separation, while the second is having a BBH progenitor at a separation with respect to a galaxy. We focus on the former in this paper because it is simpler, but the two would be related by a Poisson distribution of the rate of BBH mergers in the Universe.}. Since the positions of the galaxies are usually known to very high precision compared to that of GW sources, we assume that the uncertainty associated with them is negligible. The two-point correlation prior can be written in the Legendre polynomial basis as:

\begin{equation}
    \zeta ( \Delta_{ij}| \{C_\ell \}) = \frac{1}{4 \pi} \sum_{\ell} (2 \ell + 1)\, C_\ell \, \mathcal{P}_{\ell} (\Delta_{ij})
    \label{Eq:Galaxy_Legandre_decomposition} 
\end{equation}
Note that the prefactor here is different from Eq.~\ref{Eq:Legandre_decomposition} because here we just have just one angular integral when normalizing; over that of the BBH merger $\hat{\Omega}_i$ as compared to two angular integrals in the latter. The evidence when \hrm{data segment} $i$ is ``correlated" with galaxy $j$ is then given by,

\begin{equation}
	\begin{split}
    \mathcal{Z}^g_{ij}(\{C_\ell \}) = & \int d\vec{\lambda}_i \, \pi^0 (\vec{\lambda}_i)  \, \frac{\mathcal{L} (d_i | \vec{\lambda}_i)}{\pi^0 (\hat{\Omega}_i)} \, \zeta ( \Delta_{ij} | \{C_\ell \}), 
	\end{split}
\end{equation}
where $\vec{\lambda}_i$ are as before all the BBH parameters and $\hat{\Omega}_i$ are the directional parameters. With sample recycling this can be approximated to, 

\begin{equation}
        \mathcal{Z}^g_{ij}(\{C_\ell \}) \approx \frac{\mathcal{Z}^0_i }{M_i}\sum_n \, \frac{\zeta ( \Delta^n_{ij} | \{C_\ell \})}{\pi^0 (\hat{\Omega}^n_i)}
        \label{Eq:BBH-gal-evidence}
\end{equation}
The term $ \Delta^{n}_{ij}$ is the angular separation between the $n$-th sample in the posteriors of the $i$-th data-segment and the $j$-th galaxy. Finally using Eq.~\ref{Eq:Galaxy_Legandre_decomposition} and $\pi_o(\hat{\Omega}_i) = 1/4\pi$, the evidence becomes,

\begin{equation}
        \mathcal{Z}^g_{ij}(\{C_\ell \}) \approx \frac{\mathcal{Z}^0_i }{M_i}\sum_n \, \sum_{\ell} (2 \ell + 1) C_\ell \mathcal{P}_{\ell} (\Delta_{ij})
         \label{Eq:BBH-gal-evidence-Cl}
\end{equation}

\section{Mixture model formalism}
\label{Sec:TBS}

The expressions for two-point correlations derived in the previous sections are generally valid for measuring anisotropies with any kinds of GW data, and can be used with the various catalogues of events after accounting for selection effects as has been done with other hierarchical analyses (see for example~\cite{Stiskalek:2020wbj, Chen:2016luc, Abbott:2019yzh}). However in this section we apply them in the context of the mixture model analysis developed in Ref.~\cite{Smith:2017vfk, Smith:2020lkj}. The mixture model formalism works by applying compact binary coalescence parameter estimation on many available \hrm{data segments} without any cutoff on significance or SNR allowing us to dig deep into the background of sub-threshold events. While this removes biases due to selection effects, we need to account for the fact that only some, a priori unknown, fraction of the segments (referred to as the signal \hrm{duty cycle}) will contain a real astrophysical signal. The analysis then uses Bayesian signal and noise evidences from these \hrm{data segments} to construct posterior probability distributions for the signal \hrm{duty cycle}, as well as for the desired population \hrm{hyperparameters}. Some important details are reproduced here.

We divide the data into segments of duration $T$, chosen such that it is much larger than the inspiral \hrm{time scale} of BBH mergers in the aLIGO-aVirgo frequency band, while also being much smaller than the inverse rate of BBH mergers in the Universe. A choice of $\tau = 4 \text{s - } 16$s sits comfortably within this range. Under the assumption that there are no non-Gaussian glitches in the data, two possible hypothesis exist for each \hrm{data segment}:

\begin{enumerate}
    \item  There is a BBH signal in the \hrm{data segment}
    \item  There is only instrumental Gaussian noise in the \hrm{data segment}.
\end{enumerate}

We denote  by $\xi_S $ the signal \hrm{duty cycle}, the fraction of \hrm{data segments} which contain a BBH merger signal. With just two hypotheses, the noise \hrm{duty cycle} is then $\xi_N = 1 - \xi_S$. We then construct the mixture model likelihood for $\xi_S $ for the \hrm{data segment} $i$ using the signal evidence $\mathcal{Z}^i_S(\vec{\Lambda})$ and noise evidence $\mathcal{Z}^i_N$ for the \hrm{data segment}, 

\begin{equation}
    \mathcal{L}(d_i | \xi_S, \vec{\Lambda}) = \xi_S \, \mathcal{Z}^i_S (\vec{\Lambda}) + \xi_N \mathcal{Z}^i_N .
\end{equation}
The noise  evidence $\mathcal{Z}^i_N$ is just the likelihood that the data $d_i $ comprises only of instrumental colored Gaussian noise. The signal evidence depends on population parameters $\vec{\Lambda}$ modeled by priors $\pi(\vec{\lambda} | \vec{\Lambda})$ where $\vec{\lambda}$ are the intrinsic parameters for each event: 
\begin{equation}
    \mathcal{Z}^i_S (\vec{\Lambda}) = \int d \vec{\lambda} \, \mathcal{L}(d_i | \vec{\lambda}) \, \pi(\vec{\lambda} | \vec{\Lambda}). 
\end{equation}

\begin{figure*}[ht]
        \scalebox{0.6}{\includegraphics{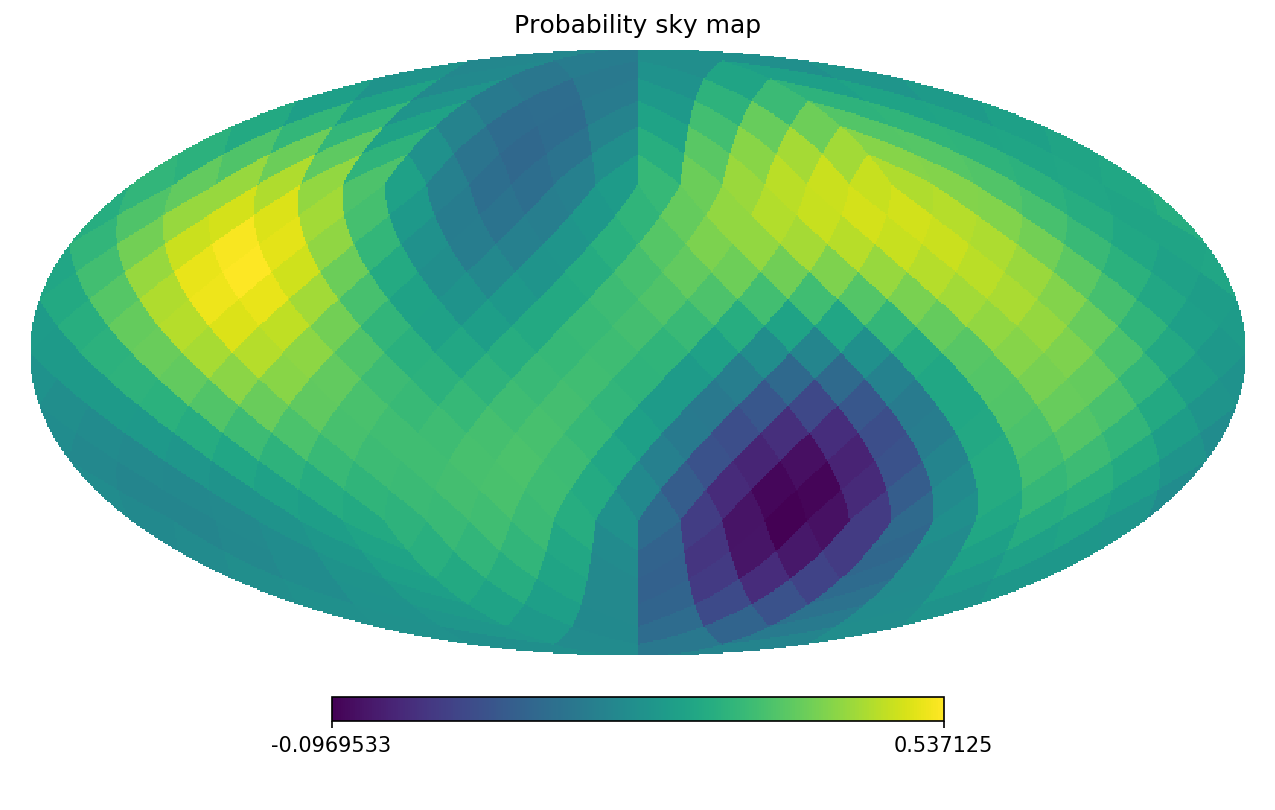}}
    
    \caption{An example Mollweide map of the probability distribution on the sky generated by using the method described in Sec.~\ref{Sec:Simulations}, with an $\ell_{max}=3$ and with $C_1=0.13, C_2=0.11 \text{ and } C_3 = 0.11$. Some of the pixels have an unphysical negative probability; no \hrm{black holes} or galaxies are allocated to those pixels in the simulations. }
    \label{fig:prob_sky_map}
\end{figure*}

With a prior for $\xi_S$ we get a posterior for the \hrm{duty cycle} $\xi_S$ and \hrm{hyperparameters} $\vec{\Lambda}$: 

\begin{equation}
  P(\xi_S, \vec{\Lambda} | d_i) = \left(\xi_S \mathcal{Z}^i_S (\vec{\Lambda}) + \xi_N \mathcal{Z}^i_N \right) \pi(\xi_S) \, \pi (\vec{\Lambda}).
    \label{Eq:base_tbs_like}
\end{equation}
 Applying this formalism to the BBH-galaxy two-point correlations is straightforward. Using $\mathcal{Z}^g_{ij} ( \{ C_{\ell} \}) $ defined in Eq.~\ref{Eq:BBH-gal-evidence-Cl}, and assuming that all the other population parameters have either been marginalized over, or are perfectly known, the posterior for correlating data segment $i$ with galaxy $j$ can be written as

\begin{equation}
	  P(\xi_S, C_{\ell} | d_i, g_j) =  \left( \xi_S \mathcal{Z}^g_{ij} ( \{C_{\ell}\} ) + \xi_N \mathcal{Z}^i_N \right) \pi(\xi_S) \pi ( \{C_\ell \} ). 
	  \label{Eq:bbh-gal-tbs}
\end{equation}
There are more hypothesis to consider when we apply the mixture model to BBH two-point correlations. For any two \hrm{data segments} $i$ and $j$ there are four hypotheses at play.

\begin{figure*}[ht]
        \scalebox{0.5}{\includegraphics{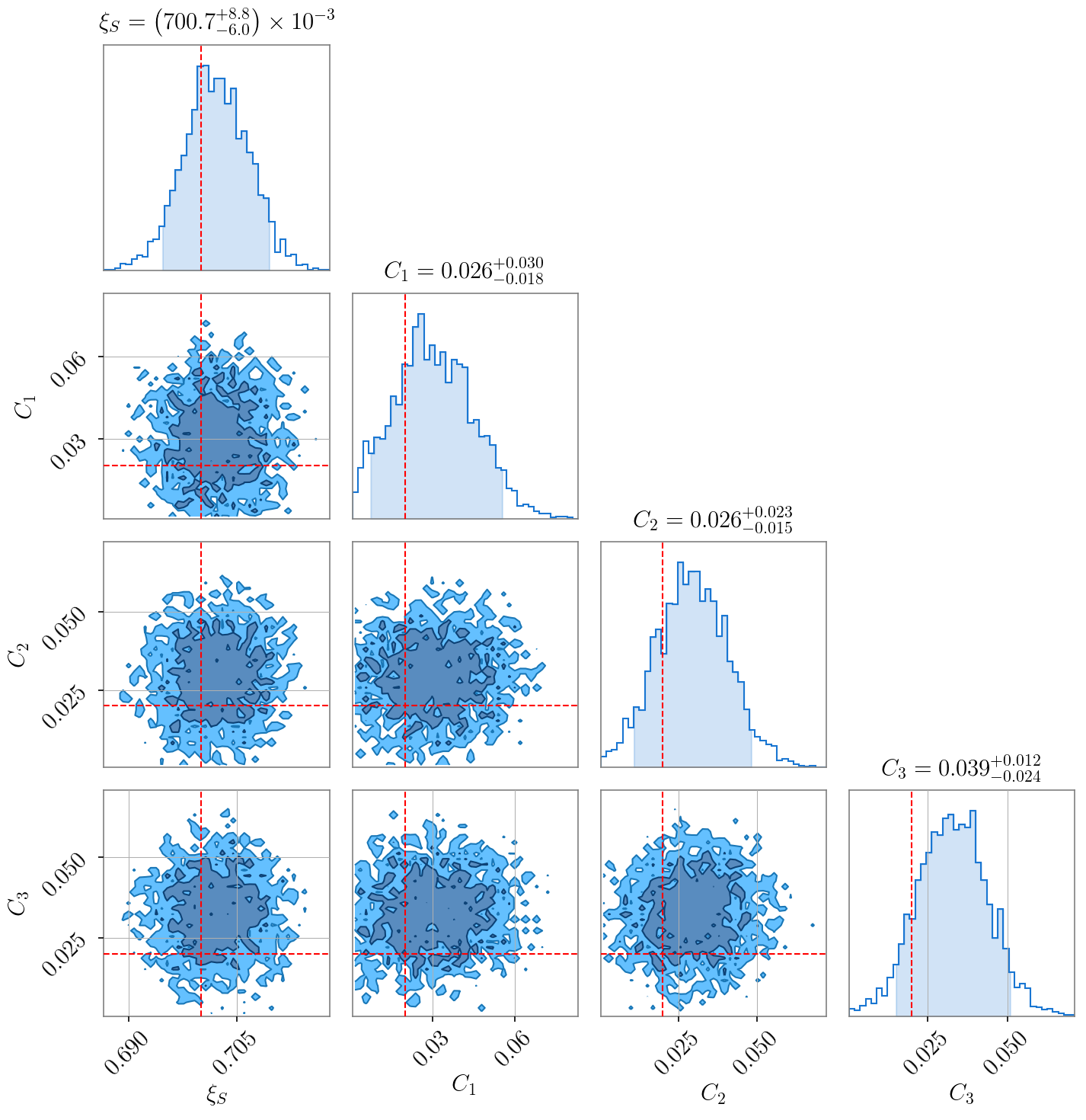}}
    
    \caption{Plot showing the posterior distributions for the angular correlations $\{C_\ell\}$, and the \hrm{duty cycle} factor $\xi_S$ for BBH-BBH two-point correlations with $2.5\times10^4$ data segments and $\ell_{max} = 3$. The monopole term is not an explicit parameter since it is normalized over and the other $\{C_{\ell}\}$ are normalized against it. The dashed red lines are the true values of the injected parameters with $\xi_S=0.7$ \redtext{ which corresponds to $17.5 \times 10^3$ BBH signals}, and $(C_1, \,	 C_2, \, C_3) = (0.018, \, 0.016, \, 0.019)$. We use uniform priors on both $\xi_S$ and $\{C_\ell \}$; 0 to 1 on the former and 0 to 0.1 on the latter. The shaded regions in the $1$-d posteriors correspond to symmetric 90\% confidence intervals. }
    \label{fig:BBH-BBH-2pt1}
\end{figure*}

\begin{enumerate}
    \item Both \hrm{data segments} have a signal: The evidence for this hypothesis is  $\mathcal{Z}^c_{ij}$ calculated in Eq.~\ref{Eq:Legdenre_2pt_decomposition} or Eq.~\ref{Eq:2pt_corr_evidence} more generally. 
    \item \hrm{Data segment} $i$ has a signal while \hrm{data segment} $j$ has only noise:  The evidence for this hypothesis is $\mathcal{Z}^0_i  \mathcal{Z}^N_j$ where  $\mathcal{Z}^0_i $ is the signal evidence calculated using the fiducial isotropic prior. 
    \item \hrm{Data segment} $j$ has a signal while \hrm{data segment} $i$ has only noise:  The evidence for this hypothesis $\mathcal{Z}^0_j  \mathcal{Z}^N_i$.
    \item Both \hrm{data segments} have only noise: The evidence for this hypothesis is $\mathcal{Z}^N_i  \mathcal{Z}^N_j$.
\end{enumerate}

The joint mixture model likelihood for correlation between GW events $i$ and $j$ is then given by, 

\begin{equation}
\begin{split}
    \mathcal{L} (d_i, d_j | \xi_S, \{C_\ell \}) = & \; \xi_S^2  \mathcal{Z}^c_{ij} (\{C_\ell \}) + \xi^2_N \mathcal{Z}^N_{j} \mathcal{Z}^N_{i} \\
     & + \,  \xi_S  \xi_N  ( \mathcal{Z}^0_{i} \mathcal{Z}^N_{j} + \mathcal{Z}^0_{j} \mathcal{Z}^N_{i} ) \end{split}
\end{equation}
Some care is needed when extending this to multiple \hrm{data segments}. Naively one might expect that two-point correlations between any two possible BBH pairs will have some extra information to be extracted. But one also needs to ensure that contradictory hypotheses are not mixed up. For example, suppose that we combine likelihoods for correlations over pairs $i-j$ and $j-k$. Then the hypothesis that both $i-j$ have a BBH merger signal is clearly incompatible with the hypothesis that both $j-k$ have only noise since they share a common \hrm{data segment}. The simplest way out of this is to multiply likelihoods only over independent pairs of \hrm{data segments}~\footnote{The number of possible pairs can be very big; with $N$ segments the number of pairs grows as $\mathcal{O}(N^2)$. We argue that any randomly chosen possible pairing is statistically valid. A heuristic argument for this is each pair-wise correlation can be thought of as a random sample from the underlying probability distribution of the correlation function. We then wish to choose a subset of the correlations to represent the distribution which is valid if the method of choosing is random, and is independent of the actual values of the correlations. It follows that any such randomly chosen set of pairs should represent the same underlying pdf to within statistical fluctuations. This is the simplest method we found, but it is possible that it does not make the optimal usage of all the information available. If a better scheme exist we leave its discovery to the future.}. The posterior for multiple segments is then:

\begin{multline}
            P (\xi_S, C_\ell | d_i, d_j ) = \prod_{i, j}\bigg [ \xi_S^2 \,  \mathcal{Z}^c_{ij} (\{C_\ell \}) \,+ \xi^2_N \mathcal{Z}^N_{j} \mathcal{Z}^N_{i}  \\ + \xi_S  \xi_N \left ( \mathcal{Z}^0_{i} \mathcal{Z}^N_{j} + \mathcal{Z}^0_{j} \mathcal{Z}^N_{i}\right )\bigg] \pi(\xi_S) \pi ( \{C_\ell \} )
    \label{Eq:TBS_for_Cl}
\end{multline}
A similar argument applies for galaxy-BBH correlations in Eq.~\ref{Eq:bbh-gal-tbs}. If we correlate a BBH merger with multiple galaxies we run the risk of multiplying contradictory hypotheses, which means we have to correlate data segments and galaxies in a one-on-one manner. Thus when extending this to multiple galaxies and data segments we again need to take products over independent pairs:

\begin{equation}
		  P(\xi_S, C_{\ell} | d_i, g_j) =  \prod_{i, j} \left( \xi_S \mathcal{Z}^g_{ij} ( \{C_{\ell}\} ) + \xi_N \mathcal{Z}^i_N \right) \pi(\xi_S) \pi ( \{C_\ell \} ). 
\end{equation}

\begin{figure*}[ht]
    \includegraphics[scale=0.55]{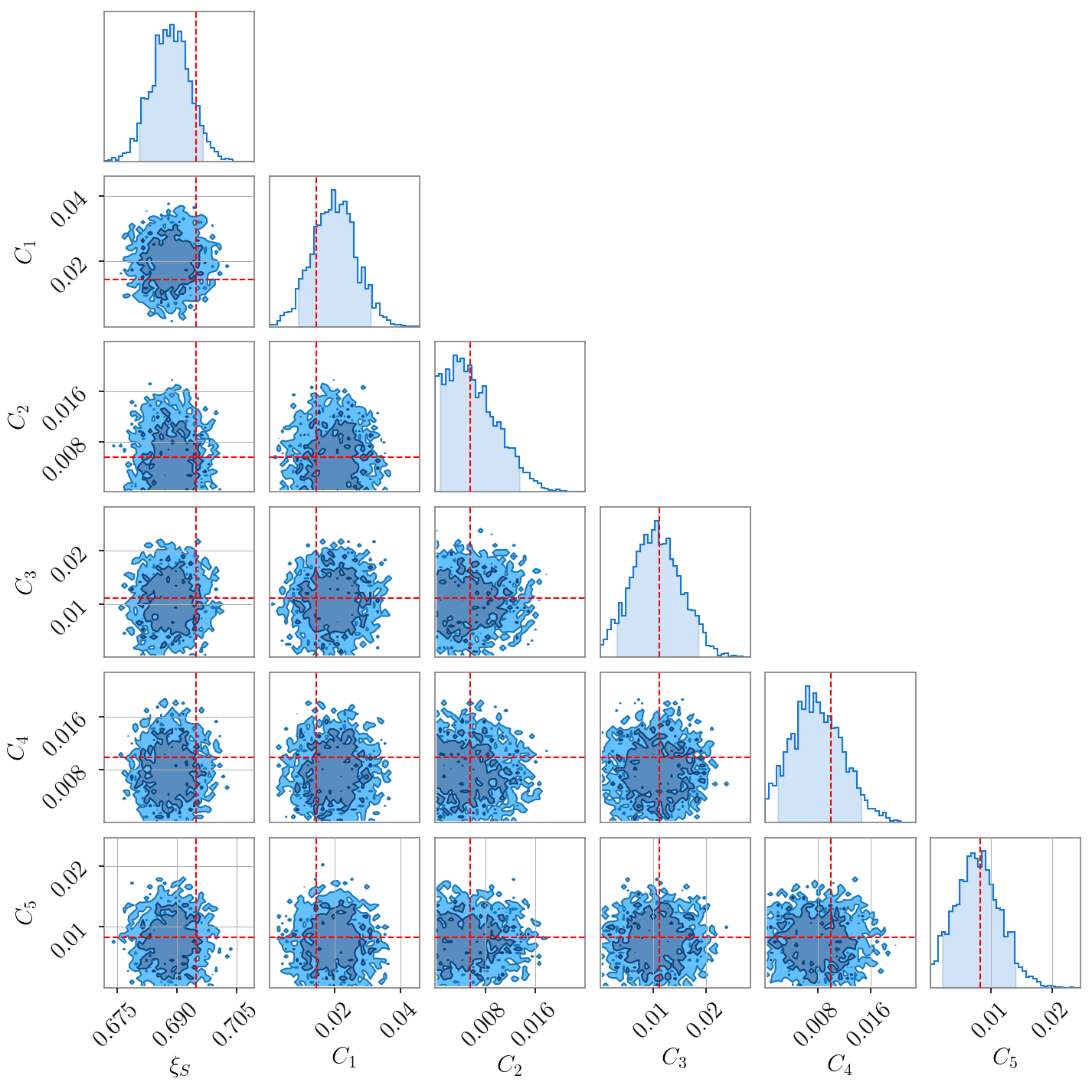}
    \caption{Plot showing the recovered $\{C_\ell\}$ as well as the \hrm{duty cycle} factor $\xi_S$ using BBH-Galaxy two-point correlations  with $2.2\times10^4$ data segments and $\ell_{max} = 5$. The monopole term is not an explicit parameter since it is normalized over and all other $C_\ell$ are normalized against it. We use uniform priors on both $\xi_S$ and $\{C_{\ell}\}$; 0 to 1 on the former and 0 to 0.1 on the latter. The shaded regions in the $1$-d posteriors correspond to symmetric 90\% confidence intervals. The dashed red lines are the true values of the injected parameters with $\xi_S=0.7$ \redtext{ which corresponds to $15.4 \times 10^3$ BBH signals} and $(C_1, \,	 C_2, \, C_3, \, C_4, \, C_5) = (0.014, \, 0.006, \, 0.011, 0.01, 0.008)$. }
    \label{fig:Gal-BBH-2pt}
\end{figure*}

\section{Simulations and Recovery}
\label{Sec:Simulations}
We simulated the GW data by first generating a large number of BBH signals using the IMRPhenomPv2 waveforms~\cite{Schmidt:2012rh, Khan:2015jqa} distributed isotropically over the sky between $0.5$ Gpc to 5 Gpc in luminosity distance uniform in comoving volume,  and in 4s segments. The signals were then added without any overlap to simulated aLIGO and aVirgo design sensitivity instrumental noise. We then ran CBC parameter estimation algorithm over each segment using the same waveform to get posteriors and evidences using fiducial isotropic angular priors. We also ran the parameter estimation algorithm over segments which contained only simulated instrumental noise. Both the simulations and the parameter estimation were done using the BILBY pipeline~\cite{Ashton:2018jfp}, with the nested sampling package DYNESTY~\cite{Speagle_2019} used for the latter. 

From this large database of segments and posteriors, we generate anisotropic simulations with desired values of $\{C_{\ell}\}$ and $\xi_S$ by probabilistically choosing segments based on the true \hrm{sky position} of the signal. To do this we pixelize the sky with Healpix~\cite{healpy1, healpy2} and calculate a probability map on the sky by drawing from a multivariate Gaussian distribution. \bluetext{We calculate the mean and the covariance matrix of the Gaussian using the chosen values of $\{C_{\ell}\}$. The monopole i.e $C_0$ gives the mean of the multivariate Gaussian. The covariance matrix can be computed by calculating he two-point correlation between pixels using the higher multipoles; dipole and above}. The probability map along with the desired signal \hrm{duty cycle} $\xi_S$ dictate the number of BBH events in each pixel, which are then randomly chosen from the previously generated database of simulated signals.

\bluetext{Fig.~\ref{fig:prob_sky_map} show an example probability map generated with this method. Since the Legendre expansion describes a real field, it is possible that some of the pixels will have negative probability values. Such pixels are excised by setting their probabilities to zero so that no BBH mergers or galaxies are assigned to them. As we go towards smaller multipole moments relative to the monopole this problem is expected to disappear. Any simulations made through $\{C_{\ell}\}$ will be susceptible to two kinds of noise. One is variance due a specific realization of the map; this is similar to cosmic variance. The second is Poisson shot noise in the pixel. In order to correct for the noise effects and the excision of pixels we compute the $\{C_{\ell}\}$ values of the maps once they are made, and use those as the true values. While this is a simplistic solution, a more sophisticated correlation function modeling the noise effects could also be used to account for them.} 

All simulations shown in this paper consist of 4s segments for aLIGO Hanford, aLIGO Livingston and aVirgo interferometers. The random fraction of segments which contain a signal is given by the duty cycle value, chosen to be $\xi_S = 0.7 $ for all simulations, with the rest being just Gaussian instrumental noise. The \hrm{duty cycle} value is chosen for computational reasons and is very large compared to realistic astrophysical rates. Instead, as a metric we will use the effective \hrm{time scales} of the simulations, defined here as the amount of real data needed to have the same number of BBHs as in the simulation assuming an average rate of 1 BBH every 4 minutes. For the $2.5 \times 10^4$, 4s long segments used for BBH two-point correlations this implies an the effective \hrm{time scale} is $\sim 48$ days with $\xi_S = 0.7 $. For the BBH-galaxy correlations with $2.2 \times 10^4$ segments  this gives a time scale of $\sim 42$ days. 

Recovery corner plots from analyzing the BBH simulations with the two-point correlation method described in Sec.~\ref{Sec:BBH-2pt} are shown in Fig.~\ref{fig:BBH-BBH-2pt1}. We assume an $\ell_{max} = 3$ which is the same value used in generating the simulation. All higher multipole moments are set to zero. \bluetext{The $\{C_{\ell}\}$ describe the statistical correlations at different angular scales. The corner plots demonstrate that the statistical properties of the background are well recovered by the methods described in this paper.} 

 For the case of BBH-Galaxy correlation, we generated simultaneous simulations of BBH signals and galaxy counts. The BBH simulations were done in the same way as before, while a simulated map of galaxy positions was made through rejection sampling using the same probability map made for the GW case. We then measure the two-point correlation function by correlating posteriors of the GW data set with a mock all-sky galaxy catalog using the methods described in Sec.~\ref{Sec:BBH-Gal-2pt}. Recovery plots from  this analysis are shown in Fig.~\ref{fig:Gal-BBH-2pt}. Since galaxies have negligible uncertainty in sky position the correlation allows us to probe deeper into the common statistical distribution of galaxies and BBH progenitors. In addition to smaller anisotropy values, the posteriors also demonstrate recovery of higher order anisotropies by successfully recovering Legendre coefficients with $\ell_{max}=5$ with a smaller amount of GW data.

\section{Application to real data}
\label{Sec:DQ}

\subsection{Glitch Hypothesis}

While this paper relies only on simulations made in stationary Gaussian data, a brief discussion of data quality is in order to access applicability to real data. As pointed out in Ref.~\cite{Smith:2017vfk}, handling non-Gaussian artifacts in GW detectors (called glitches) requires us to introduce additional hypotheses for each segment. A conservative assumption is used that glitches look like single detector BBH signals. The Bayesian evidence that there is a glitch in a segment in detector $1$ is then just the evidence for a single detector signal hypothesis i.e,

\begin{equation}
	Z_{(1)}^g \equiv Z_{(1)}^S  = \int d\vec{\lambda}_{(1)} P(\vec{\lambda})_{(1)})\, \pi(\vec{\lambda}_{(1)}). 
 \end{equation}
The subscript here is an index over detectors while $\vec{\lambda}$ consist of all the BBH parameters as before. In Ref.~\cite{Smith:2017vfk} individual glitch hypotheses are constructed for each detector and are used to measure the glitch \hrm{duty cycles} for each individual detector. We simplify that somewhat by constructing a single catch-all hypothesis that there is a glitch at any one of the detectors in a data segment. For this hypothesis we rely on the assumption that it is unlikely for a glitch to occur along with a signal in a segment, and that it is also unlikely for glitches to occur in two or more detectors in the same segment. Under these assumptions the glitch evidence for a segment for the case of three detectors is:

\begin{equation}
\mathcal{Z}^g = Z_{(1)}^N Z_{(2)}^N Z^S_{(3)} +  Z_{(3)}^N Z_{(1)}^N Z^S_{(2)} +  Z_{(2)}^N Z_{(3)}^N Z^S_{(1)} 
\label{Eq:Glitch_hypothesis}
\end{equation}
The equivalent of Eq.~\ref{Eq:TBS_for_Cl} then becomes

\begin{equation}
\begin{split}
    \mathcal{L} (d_i, d_j | \xi_S, \xi_g, \{C_\ell \}) = & \prod_{i, j}\bigg [ \xi_S^2 \, \mathcal{Z}^c_{ij} (\{C_\ell \}) + + \xi^2_N \mathcal{Z}^N_{j} \mathcal{Z}^N_{i} + \\ & \xi_S  \xi_N \left ( \mathcal{Z}^0_{i} \mathcal{Z}^N_{j} + \mathcal{Z}^0_{j} \mathcal{Z}^N_{i}\right )  + \\ & \xi_g \xi_N  \left ( \mathcal{Z}^g_{i} \mathcal{Z}^N_{j} + \mathcal{Z}^g_{j} \mathcal{Z}^N_{i}\right ) + \\ & \xi_g \xi_S  \left ( \mathcal{Z}^g_{i} \mathcal{Z}^0_{j} + \mathcal{Z}^g_{j} \mathcal{Z}^0_{i}\right ) \bigg],
\end{split}
\end{equation}
where $\xi_g $ is the glitch \hrm{duty cycle} i.e the fraction of segments containing a glitch in one of the detectors. The \hrm{duty cycle} factors are now related as $\xi_S  + \xi_N  + \xi_g = 1$. We note again that this depends on coincident glitches between detectors being unlikely and additional data quality cuts might be required to ensure this requirement is met with real data. We will defer application of the N-point correlation methods to real data to a future work. 

\subsection{Application to real galaxy catalogs}

The simplistic galaxy simulation in this paper assumes that we can measure the galaxy field across the entire sky with equal sensitivity, which is not true for real galaxy surveys especially because of obstruction from the dust and gas of the Milky Way. This effect is usually modeled by assuming that the observed field is filtered through a window function which captures the incompleteness of the observed galaxy distribution across the sky and has the effect of changing the spherical harmonic (and hence the multipole) expansion of the galaxy distribution (see for eg~\cite{Dodelson:2003}). This would need to be accounted for when correlating with a real galaxy catalog. 

We also point out that the two-point cross-correlation evidence described in Eq.~\ref{Eq:BBH-gal-evidence} assumes a one-to-one pairing between GW data segments and galaxies. Since reusing them is not possible, we are forced  to have the same number of galaxies as we have segments. Real galaxy catalogs will of course have tens or hundreds of millions of galaxies at the very least. One way to apply this formalism to \hrm{cross correlating} with real catalogs would be to randomly sample from them. For example if one is working with a million GW data segments one can randomly pick a million galaxies from the a catalog like SDSS and correlate them one on one with the GW data segments.

\subsection{Sensitivity}

\redtext{While the broad localization of events in the GW posterior distributions is a major source of uncertainty for detecting and measuring anisotropies, another important source of noise arises from the Poisson statistics of the events. Under this shot noise, we would expect the uncertainty in measurements of spherical harmonic coefficients $a_{\ell m}$'s to fall as $1/\sqrt{N}$, where $N$ are the number of events. Likewise uncertainty in measurements of $C_\ell$'s should scale as as $1/N$. If the two-point correlation method is applied to a catalog of GW events, then to measure a dipole anisotropy of $C_1 \sim 0.01$ we would need $\mathcal{O}(100)$ events to overcome the shot noise floor. This is broadly consistent with a simulated analysis done by Ref.~\cite{Payne:2020pmc}, albeit in a context of next generation detectors. }

\redtext{When the two-point correlation analysis is applied to sub-threshold events as is done in this paper, predicting sensitivity becomes more complicated. While the shot noise remains unchanged, an accurate estimate of the sensitivity would need simulations based on astrophysical realistic duty-cycles and population distributions, along with glitch rates of the detectors. For a distance cut off of $5 \, Gpc$ and with a realistic duty cycle of $4 \times 10^{-4}$, the required time of detection of an isotropic signal was estimated to be $\approx 20$ hours in~\cite{Smith:2017vfk}. Assuming the same shot noise based scaling as above, we can then estimate that it would take $\mathcal{O}(100)$ days of data to detect $C_1 \sim 0.01$.}

\section{Conclusion}

In this paper we have developed ways to measure the statistical $N$-point correlations of the angular distribution of BBH mergers, with emphasis on the specific case of two-point correlations. We have also shown how the two-point correlation method can be used to \hrm{cross correlate} BBH distribution with other tracers of large scale structure. Using the formalism developed in~\cite{Smith:2017vfk} and~\cite{Smith:2020lkj} we have demonstrated measurement of anisotropies on simulated data using two point correlations. This method holds promise to delve deeper into the noise floor than standard stochastic searches and to measure anisotropies in the ensemble of binary mergers. The formalism can be extended to measure higher order multipoles too if so desired. 

Recently there have been studies on correlating \redtext{GW} data with the distribution of galaxies~\cite{Mukherjee:2018ebj, Mukherjee:2019oma, Mukherjee:2019wcg}, and in particular on correlating the anisotropic stochastic maps from aLIGO-aVirgo with galaxy counts~\cite{sgwb_sdss_cc}. The methods developed in this paper could provide a boost to such efforts. Theoretical modeling of the stochastic background from stellar mergers also suggests that BBH-galaxy correlations would be less susceptible to Poisson noise when measuring anisotropies than the \redtext{GW} side alone due to the relatively small number of BBH events~\cite{Canas-Herrera:2019npr, Alonso:2020mva}. Finally, galaxy-BBH correlation could in principle allow us to probe differences in the relative distribution of galaxies and progenitors of GW. But this would perhaps require angular resolutions much smaller than possible with the current generation of detectors.

There are several ways to extend or apply the formalism developed in this work. \redtext{One can apply the angular two-point correlation method to the catalog of events already published accounting for selection effects}. The Bayesian posteriors also let us access the distance measurements of the events, so one can also consider measuring correlations in three dimensions rather than just over the two sphere. This would give us the ability to directly measure the three dimensional structure of matter, and constrain the \hrm{power spectrum} of BBH progenitors though GW. The binary neutron star merger GW170817 demonstrated an application of GW towards cosmology through a GW measurement of the Hubble constant~\cite{Abbott:2017xzu}. A similar idea was also recently explored in Ref.~\cite{Vijaykumar:2020pzn} in the context of third generation detectors assuming a Gaussian localization of sources. Measurement of the \hrm{power spectrum} of matter distribution will enlarge the scope of GW as a tool for cosmological inference. \redtext{The extension to three-dimensional two-point correlations can also be applied to catalogues of events to probe local structure. We will explore these ideas further in a future paper, with application to GW catalogs and assessing the sensitivity of both the current generation of detectors and with the next generation which will have deeper redshift reach and more precise localization.} 
\\

\section{Acknowledgments}
We are grateful to Andrew Matas and Colm Talbot for useful discussion and comments. S.B acknowledges support by the Doctoral Dissertation Fellowship at the University of Minnesota. SB and VM were supported by NSF grant PHY-1806630. All corner plots were made with ChainConsumer~\cite{ChainConsumer}. The authors are thankful for the computing resources provided by LIGO Laboratory and supported by the National Science Foundation grants PHY--0757058 and PHY--0823459. This paper carries the internal LIGO document number \dcc.

\bibliography{atbs}

\end{document}